\documentclass[reprint,prl,twocolumn,showpacs]{revtex4-1}
\usepackage{txfonts}
\usepackage{float}
\usepackage{graphicx}
\begin{document}
	\title{Incoherent-coherent crossover and the pseudogap in Te-annealed superconducting Fe$_{1+y}$Te$_{1-x}$Se$_{x}$ revealed by magnetotransport measurements}
	\date{\today}
	\author{Takumi Otsuka$^1$}
	\author{Shotaro Hagisawa$^1$}
	\author{Yuta Koshika$^1$}
	\author{Shintaro Adachi$^1$}
	\author{Tomohiro Usui$^1$}
	\author{Nae Sasaki$^1$}
	\author{Seya Sasaki$^1$}
	\author{Shunpei Yamaguchi$^1$}
	\author{Yoshiki Nakanishi$^2$}
	\author{Masahito Yoshizawa$^2$}
	\author{Shojiro Kimura$^3$}
	\author{Takao Watanabe$^1$$^\star$\thanks{e-mail: twatana@hirosaki-u.ac.jp}}
	\affiliation{Graduate School of Science and Technology, Hirosaki University, 3 Bunkyo, Hirosaki, 036-8561 Japan$^1$}
	\affiliation{Graduate School of Engineering, Iwate University, Morioka 020-8551, Japan$^2$}
	\affiliation{Institute for Materials Research, Tohoku University, 2-1-1 Katahira, Aoba-ku, Sendai, 980-8577 Japan$^3$}
	\begin{abstract}
		In this study, we conducted various magnetotransport measurements on Fe$_{1+y}$Te$_{1-x}$Se$_{x}$ single crystals from which excess iron was sufficiently removed. Our results revealed that crossover from the incoherent to the coherent electronic state and opening of the pseudogap occur at high temperatures ($\approx$ 150 K for $x$ = 0.2). This is accompanied by a more substantial pseudogap and the emergence of a phase with a multi-band nature at lower temperatures (below $\approx$ 50 K for $x$ = 0.2) before superconductivity sets in. A comparison of these results with those of the as-grown (non-superconducting) samples implies that the coherent state accompanied by the pseudogap is needed for the occurrence of superconductivity in this system.      
	\end{abstract}
	\maketitle
	\section{I. INTRODUCTION}

	
	Knowledge of the doping($x$)-temperature ($T$) phase diagram of superconductors with a high superconducting transition temperature, $T_c$, and with strongly correlated electrons, such as the cuprates and Fe-based superconductors, is well recognized to be crucial to understand the mechanism of high-$T_c$ superconductivity~\cite{uchida}. Especially, the anomalous non-Fermi-liquid-like normal-state transport properties are central problems that need to be solved. Among the Fe-based superconductors, the iron chalcogenide Fe$_{1+y}$Te$_{1-x}$Se$_{x}$~\cite{fang} is unique in that its crystal structure is the simplest; it only consists of conducting Fe\textit{X} (\textit{X}:Te or Se) layers, and the electron correlation level is considered to be the strongest~\cite{yin}. Therefore, it is important to investigate the phase diagram of this system. However, the existence of excess iron (represented as $y$ in the above-mentioned formula) has thus far prevented the establishment of the true phase diagram of this system. Recently, an O$_{2}$-annealing technique was developed to remove excess iron from the Fe$_{1+y}$Te$_{1-x}$Se$_{x}$ system; hence, this technique was employed to investigate its phase diagram ~\cite{tamegai}. This pioneering work showed that the antiferromagnetic (AFM) ordered phase exists only for x $\textless$ 0.05 and bulk superconductivity emerges from x = 0.05, and that these two phases do not coexist with each other. Furthermore, the temperature dependence of the Hall coefficient, $R_{H}$, showed a peak at $T^{*}$. This result suggests that a phase of a multi-band nature appears below $T^{*}$. Presently, however, this anomalous behavior is not yet fully understood.
	
	On the other hand, the evolution from incoherent to coherent electronic states with increasing Se doping, i.e., by increasing x, in this Fe$_{1+y}$Te$_{1-x}$Se$_{x}$ system was observed by angle-resolved photoemission spectroscopy (ARPES)~\cite{ieki,shen}. That is, a broad ARPES spectrum for both the hole and electron bands near $\Gamma$ and the $M$ point, respectively, for samples with x $\le$ 0.2, which is indicative of incoherent electronic states, has been shown to progressively change to sharper ones, thereby indicating coherent electronic states, for the samples with x $\ge$ 0.4. Because the samples with x $\le$ 0.2 were found to be non-superconducting~\cite{ieki}, whereas those with x $\ge$ 0.4 were superconducting, a close relationship between the coherent electronic state and the emergence of superconductivity has been suggested. Unfortunately, however, an amount of excess iron may have been incorporated in these ARPES samples~\cite{ieki,shen}, thus the effects of excess iron and doping (i.e., the value of x) were not resolved in this previous work. Therefore, examining this crossover phenomenon more systematically and establishing the phase diagram is expected to be a great challenge.
	
\begin{figure}[t]
		\begin{center}
			\includegraphics[width=85mm]{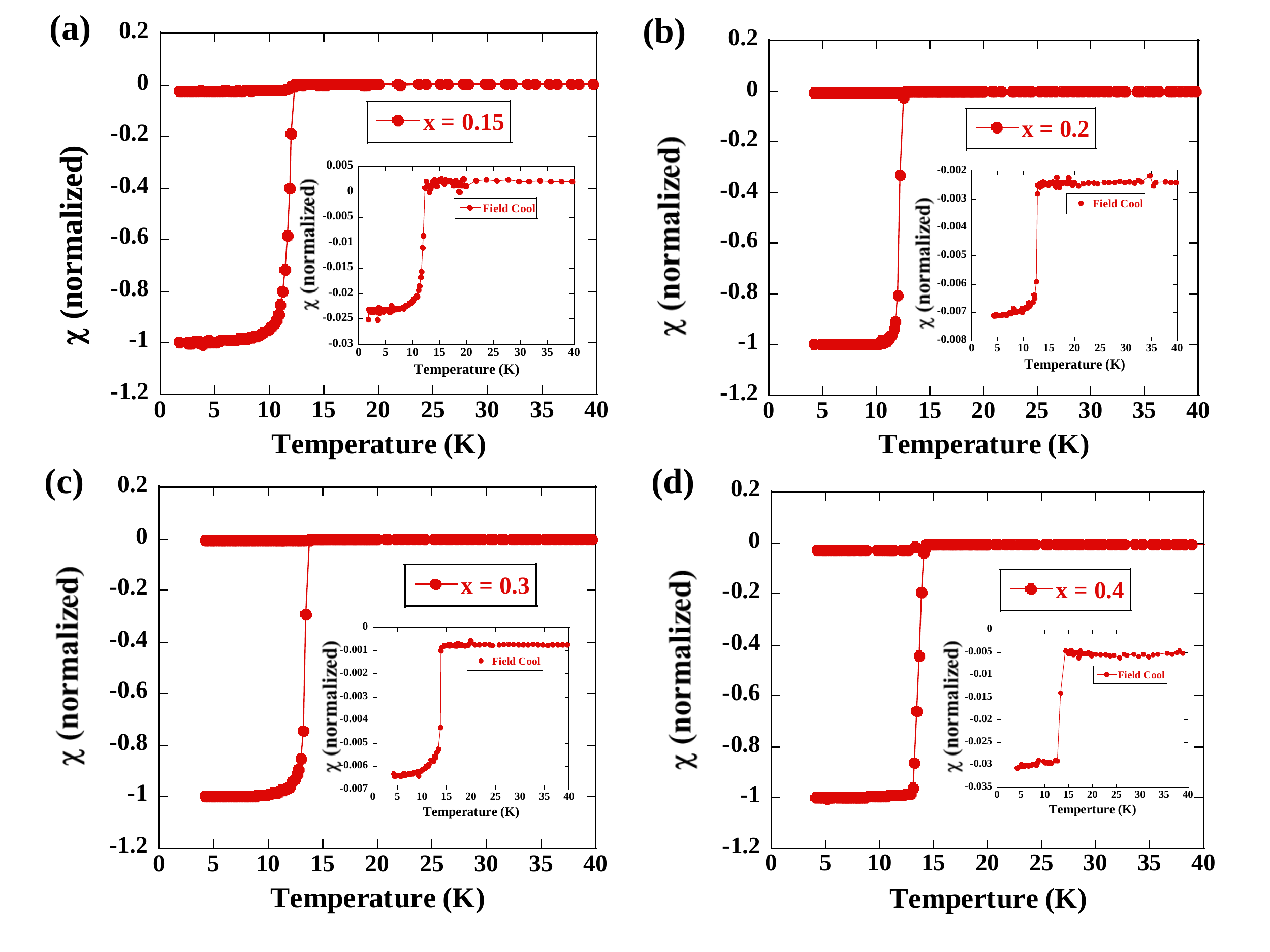}
			\caption{\label{figS1}(Color online)  Temperature dependence of the zero-field-cooled (ZFC) normalized magnetic susceptibilities $\chi$ for Te-annealed Fe$_{1+y}$Te$_{1-x}$Se$_{x}$ single crystals with (a) $x$ = 0.15, (b) $x$ = 0.2, (c) $x$ = 0.3, and (d) $x$ = 0.4. The insets show the field-cooled (FC) data. Data were recorded by applying a magnetic field of 30 Oe parallel to the $c$-axis.}
		\end{center}
	\end{figure}

	Here, we address this issue by conducting transport measurements such as determining the Hall coefficient $R_{H}$ and magnetoresistance (MR), as well as the in-plane and out-of-plane resistivities, $\rho_{ab}$ and $\rho_{c}$, respectively, at zero field, focusing on the sample with low Se concentration x = 0.2. Prior to this study, we developed a new annealing method to remove excess iron (hereafter denoted as ``Te-anneal"), in which single crystals are annealed under tellurium vapor~\cite{koshika}. This Te-annealing method has the advantage of sufficiently removing the excess iron without damaging the samples, even those with lower Se concentrations (it is known that the smaller the amount of doping x, the more difficult it is to remove excess iron). Because the as-grown samples are non-superconducting (superconductivity is filamentary even if it exists~\cite{koshika}) and the annealed samples are fully superconducting (Fig. \ref{figS1}), this Te-annealing method has provided us with the unique opportunity to study the transport properties of both types of samples (non-superconducting and superconducting) comparatively for a fixed amount of doping x. Furthermore, Te-annealing has enabled us to measure $\rho_{c}$, because we can obtain thick samples along the c-axis even after annealing. We find that the annealed samples show anomalies in $\rho_{ab}$, $\rho_{c}$, and $R_{H}$ at $T^*_{\rho_{ab}}$, $T^*_{\rho_{c}}$, $T^{**}_{R_{H}}$, and $T^*_{R_{H}}$, respectively, whereas the as-grown sample does not exhibit these anomalies. Furthermore, we observe markedly negative MR for both the samples. Based on these observations, we decided to perform a simple two-band analysis. The results show that only the superconducting samples undergo a crossover from incoherent to coherent, accompanied by the opening of a pseudogap with a broad boundary at  $T^{**}_{R_{H}}$ and $T^*_{\rho_{ab}}$ ($\approx$ 150 K). Subsequently, a phase with a multi-band nature appears below $T^*_{\rho_{c}}$ and $T^*_{R_{H}}$ ($\approx$ 50 K), implying that the crossover plays an important role in the occurrence of superconductivity in this FeTe$_{1-x}$Se$_{x}$ system.
	
	\section{II. EXPERIMENT}
	Single crystals of Fe$_{1+y}$Te$_{1-x}$Se$_{x}$ (0$\le x \le$0.4) were grown using the Bridgman method~\cite{koshika}. The nominal composition of Fe was set to 1.03, which is in common with that of the Se-doped samples. For Fe$_{1+y}$Te, it was set to 1.13. Selected as-grown crystals were cleaved into smaller crystals $\approx$ 1 mm thick, and they were annealed under tellurium vapor (Te-anneal)~\cite{koshika} for more than 400 h at 400 $^\circ$C. Electron probe microanalysis (EPMA) showed that the amount of excess iron, $y$, in the annealed samples was roughly zero, whereas the amount of doping $x$ was unchanged. Magnetic susceptibility measurements were performed using a superconducting quantum interference device (SQUID) magnetometer (Quantum Design MPMS) with a magnetic field of 30 Oe applied parallel to the $c$-axis.
	
	The in-plane resistivity $\rho_{ab}$ was measured using the standard DC four-terminal method. The out-of-plane resistivity $\rho_{c}$ was measured using the modified DC four-terminal method, in which voltage contacts were attached to the center of the {\it ab} plane, and the current contacts almost covered the entire remaining space~\cite{motohashi}. The Hall resistivity $\rho_{yx}$ and magnetoresistance (MR) were simultaneously measured using a Physical Property Measurement System (Quantum Design PPMS) with the five-terminal method with the applied field parallel to the $c$-axis. The value of $\rho_{yx}$ was obtained by averaging the difference of the data set at positive and negative fields, i.e., $\rho_{yx}(H) = (\rho_{yx}(+H) - \rho_{yx}(-H))/2$, which can eliminate the MR component due to the misalignment of contacts.   	
	
	\begin{figure*}[t]
		\begin{center}
			\includegraphics[width=180mm]{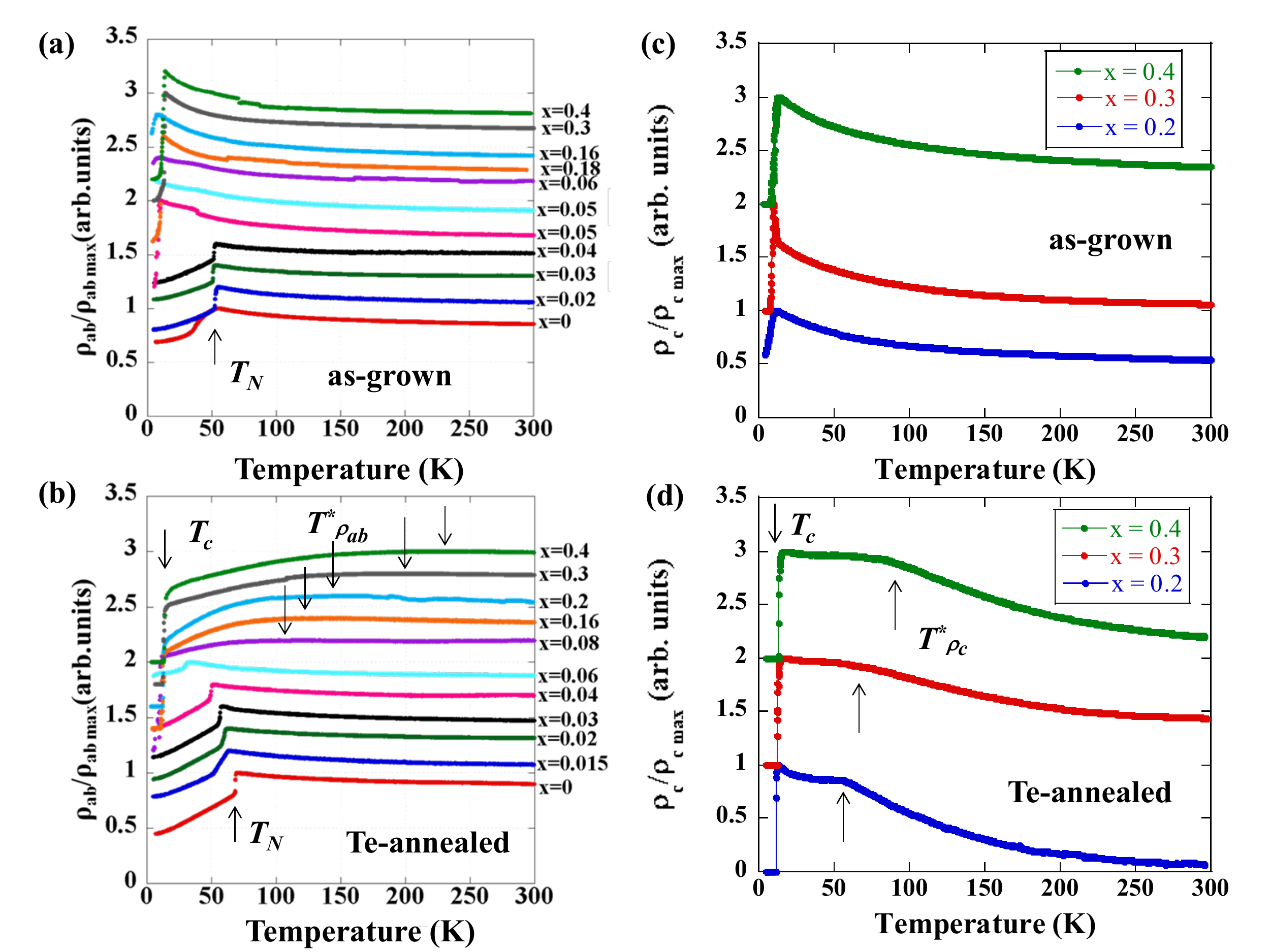}
			\caption{\label{fig1}(Color online)  Normalized in-plane resistivity $\rho_{ab}(T)$ for the (a) as-grown and (b) Te-annealed Fe$_{1+y}$Te$_{1-x}$Se$_{x}$ (0 $\le$ x $\le$ 0.4) single crystals, and normalized out-of--plane resistivity $\rho_{c}(T)$ for the (c) as-grown and (d) Te-annealed Fe$_{1+y}$Te$_{1-x}$Se$_{x}$ (0.2 $\le$ x $\le$ 0.4) single crystals. The resistivity data are shifted vertically for clarity. The arrows indicate the superconducting, AFM transitions, and characteristic temperatures, $T^*_{\rho_{ab}}$ and $T^*_{\rho_{c}}$.}   
		\end{center}
	\end{figure*}

	\begin{figure}[t]
		\begin{center}
			\includegraphics[width=85mm]{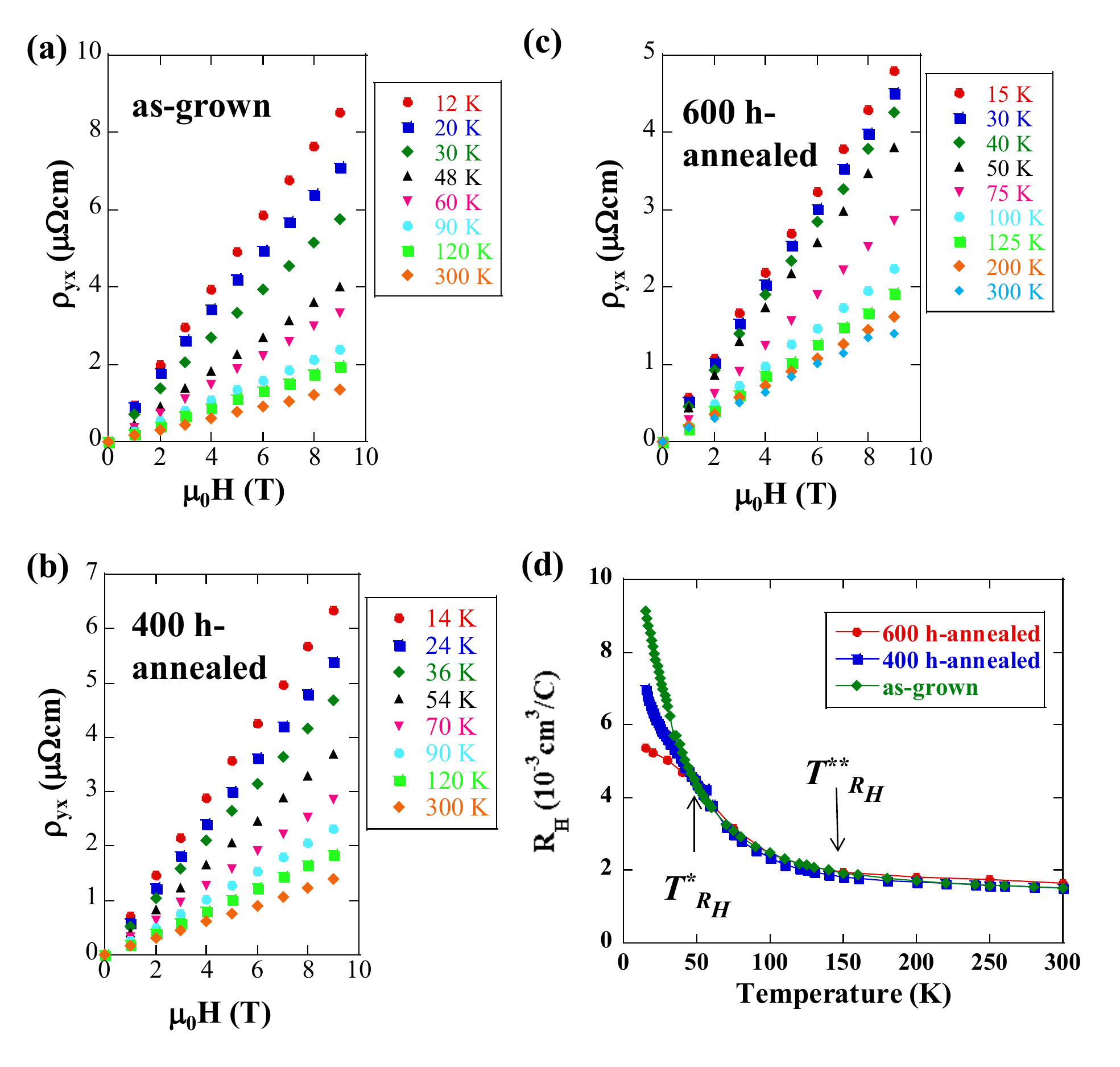}
			\caption{\label{fig2}(Color online) Hall resistivity, $\rho_{yx}$, at several temperatures for the (a) as-grown, (b) 400 h-annealed and (c) 600 h-annealed Fe$_{1+y}$Te$_{0.8}$Se$_{0.2}$ single crystals. (d) Temperature dependence of Hall coefficients, $R_{H}$, for the as-grown, 400 h-annealed and 600 h-annealed Fe$_{1+y}$Te$_{0.8}$Se$_{0.2}$ single crystals. The arrow indicates the characteristic temperature $T^*_{R_{H}}$.}
		\end{center}
	\end{figure}

	\section{III. RESULTS}
	Figure \ref{figS1} (a)--(d) show the temperature dependence of the magnetic susceptibilities $\chi$ of the Te-annealed Fe$_{1+y}$Te$_{1-x}$Se$_{x}$ single crystals, for $x$ = 0.15, 0.2, 0.3, and  0.4, respectively. The main panels show the zero-field-cooled (ZFC) data, whereas the insets show the field-cooled (FC) data. All the plotted data show sharp superconducting transitions ($\Delta$ $T_{c}$ $\le$ 1 K) with the onset temperatures of 12.1, 12.8, 13.7, and 14.5 K, for $x$ = 0.15, 0.2, 0.3, and 0.4, respectively. Here, the superconducting transition width $\Delta$ $T_{c}$ was determined by the interval between the onset temperature of the ZFC signal and the temperature at which this signal reached 90 $\%$ of its maximum value. These data confirm that the superconducting properties of the Te-annealed samples are very good. In addition, it should be remarked that all the samples show the Meissner signal (negative value in the FC data) even though the signal is very weak, which may be attributed to the sufficient removal of excess iron by Te-annealing.
	
	Figure \ref{fig1}(a) and (b) show $\rho_{ab}(T)$ for the as-grown and fully Te-annealed Fe$_{1+y}$Te$_{1-x}$Se$_{x}$ (0$\le$x$\le$0.4) single crystals, respectively. These data reproduce the overall features of the O$_2$ annealed crystals~\cite{tamegai}. The as-grown crystals show a resistivity drop that originated in the long-range AFM transition at $\approx$ 50 K in the doping region of 0$\le$x$\le$0.04, whereas for $x \ge$0.05, they show filamentary (rather than bulk in nature) superconductivity~\cite{koshika}. For the annealed crystals, the AFM transition is observed at $T_{N} \approx$ 70 K for $x$ = 0, after which it decreases with increasing x to $\approx$ 30 K for x = 0.06, and bulk superconductivity appears for $x$ $\ge$ 0.08~\cite{koshika}. It should be noted that the temperature dependence of $\rho_{ab}$ for all the superconducting crystals appears as a poorly resolved broad structure. Here, the temperature at which $\rho_{ab}$ reaches its maximum is defined as $T^*_{\rho_{ab}}$. $T^*_{\rho_{ab}}$ linearly increases from 110 K for x = 0.08 to 230 K for x = 0.4. 
	
	Figure \ref{fig1}(c) and (d) show $\rho_{c}(T)$ for the as-grown and fully Te-annealed Fe$_{1+y}$Te$_{1-x}$Se$_{x}$ (x = 0.2, 0.3, and 0.4) single crystals, respectively. The as-grown crystals show smooth semiconducting behavior before the filamentary superconducting transition. On the other hand, the annealed crystals show semiconducting behavior at higher temperatures similar to the as-grown crystals; however, they show the typical plateau below the temperatures, $T^*_{\rho_{c}}$, before the bulk superconducting transition. $T^*_{\rho_{c}}$ is estimated to be $\approx$ 55, $\approx$ 65, and $\approx$ 90 K for x = 0.2, 0.3, and 0.4, respectively. Here, $T^*_{\rho_{c}}$ is estimated as the temperature at which the second-order derivative is minimized. Although this plateau appears highly anomalous, our observation is consistent with a previous report about an O$_2$ annealed x = 0.4 crystal~\cite{noji1}.
	
	We examined the changes in the electronic system as a result of Te-annealing by measuring the temperature dependence of the Hall coefficients, $R_{H}$, for the Fe$_{1+y}$Te$_{0.8}$Se$_{0.2}$ single crystals. Hereafter, our discussion concentrates on the measurements of the x = 0.2 crystals. Figure \ref{fig2}(a)-\ref{fig2}(c) show the magnetic field dependence of the Hall resistivity, $\rho_{yx}$, at several temperatures for the as-grown, 400 h-annealed, and 600 h-annealed Fe$_{1+y}$Te$_{0.8}$Se$_{0.2}$ single crystals, respectively. In all cases, $\rho_{yx}$ increases linearly with the applied magnetic fields maintaining a positive slope, $d\rho_{yx}/dH \textgreater 0$, down to low temperatures slightly above $T_{c}$. These results indicate that the hole-type carrier dominates the electron transport. It should be noted, however, that this field-linear behavior in $\rho_{yx}$ is quite contrary to that of the FeTe$_{0.5}$Se$_{0.5}$ thin films\cite{tsukada} and Fe$_{1+y}$Te$_{0.6}$Se$_{0.4}$ single crystals~\cite{sun}, for which $\rho_{yx}$ showed marked nonlinear behavior when exposed to magnetic fields below $\approx$ 40 K. We suppose that some kind of carrier imbalance between the electrons and holes occurs for the samples with $x$ $\ge$ 0.4.  
	
	\begin{figure}[t]
		\begin{center}
			\includegraphics[width=85mm]{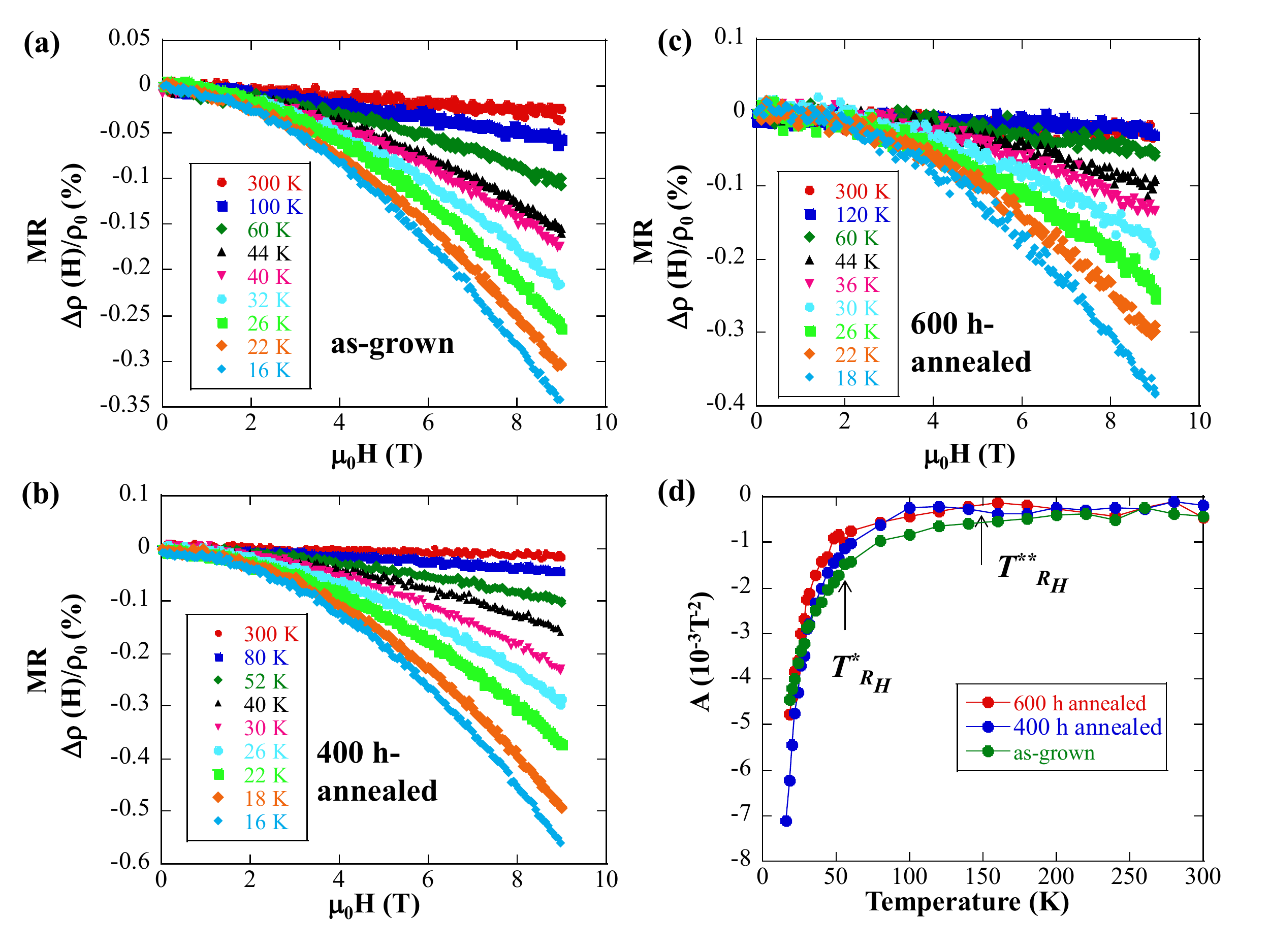}
			\caption{\label{fig3}(Color online)
				Magnetoresistance (MR), at several temperatures for the (a) as-grown, (b) 400 h-annealed and (c) 600 h-annealed Fe$_{1+y}$Te$_{0.8}$Se$_{0.2}$ single crystals. (d) Temperature dependence of coefficient A for the as-grown, 400 h-annealed, and 600 h-annealed Fe$_{1+y}$Te$_{0.8}$Se$_{0.2}$ single crystals. The arrows indicate the characteristic temperatures, $T^{**}_{R_{H}}$ and $T^*_{R_{H}}$.}
		\end{center}
	\end{figure}
	
	\begin{figure*}[t]
		\begin{center}
			\includegraphics[width=180mm]{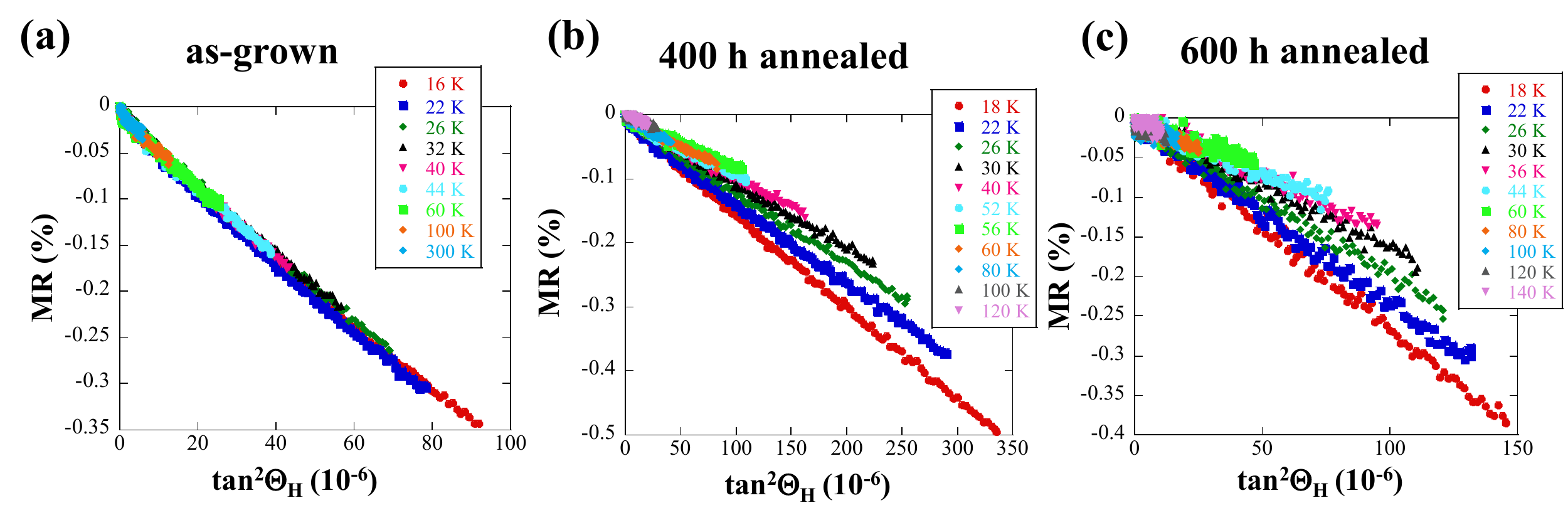}
			\caption{\label{figS2}(Color online)  Magnetoresistance (MR) as a function of tan$^2\Theta_{H}$ at several temperatures for the (a) as-grown, (b) 400 h-annealed, and (c) 600 h-annealed Fe$_{1+y}$Te$_{0.8}$Se$_{0.2}$ single crystals.}   
		\end{center}
	\end{figure*}
	
	Then, $R_{H} = \rho_{yx}/\mu_{0}H$ for the as-grown, 400 h-annealed, and 600 h-annealed crystals were plotted as a function of temperature and are shown in Fig. \ref{fig2}(d). All the samples show similar behavior from room temperature to $\approx$ 50 K: $R_{H}$ has an almost constant positive value from room temperature to $\approx$ 150 K, and below this temperature it gradually increases with decreasing temperature. On the other hand, below 50 K, each of the samples was observed to exhibit different behavior. $R_{H}$ for the as-grown crystal rapidly increases with decreasing temperature, whereas for the 400 h-annealed crystal, the increasing trend weakens, and for the 600 h-annealed crystal it stabilizes to show rather plateau-like behavior. This result agrees with that in the previous report\cite{tamegai} and suggests that the multi-band nature manifests itself below 50 K. The temperatures at which $R_{H}$ starts to increase ($\approx$ 150 K) and ramifies ($\approx$ 50 K) are defined as $T^{**}_{R_{H}}$ and $T^*_{R_{H}}$, respectively. Remarkably, $T^{**}_{R_{H}}$ and $T^*_{R_{H}}$ coincide with $T^*_{\rho_{ab}}$ and $T^*_{\rho_{c}}$, respectively.

	\begin{figure}[t]
		\begin{center}
			\includegraphics[width=85mm]{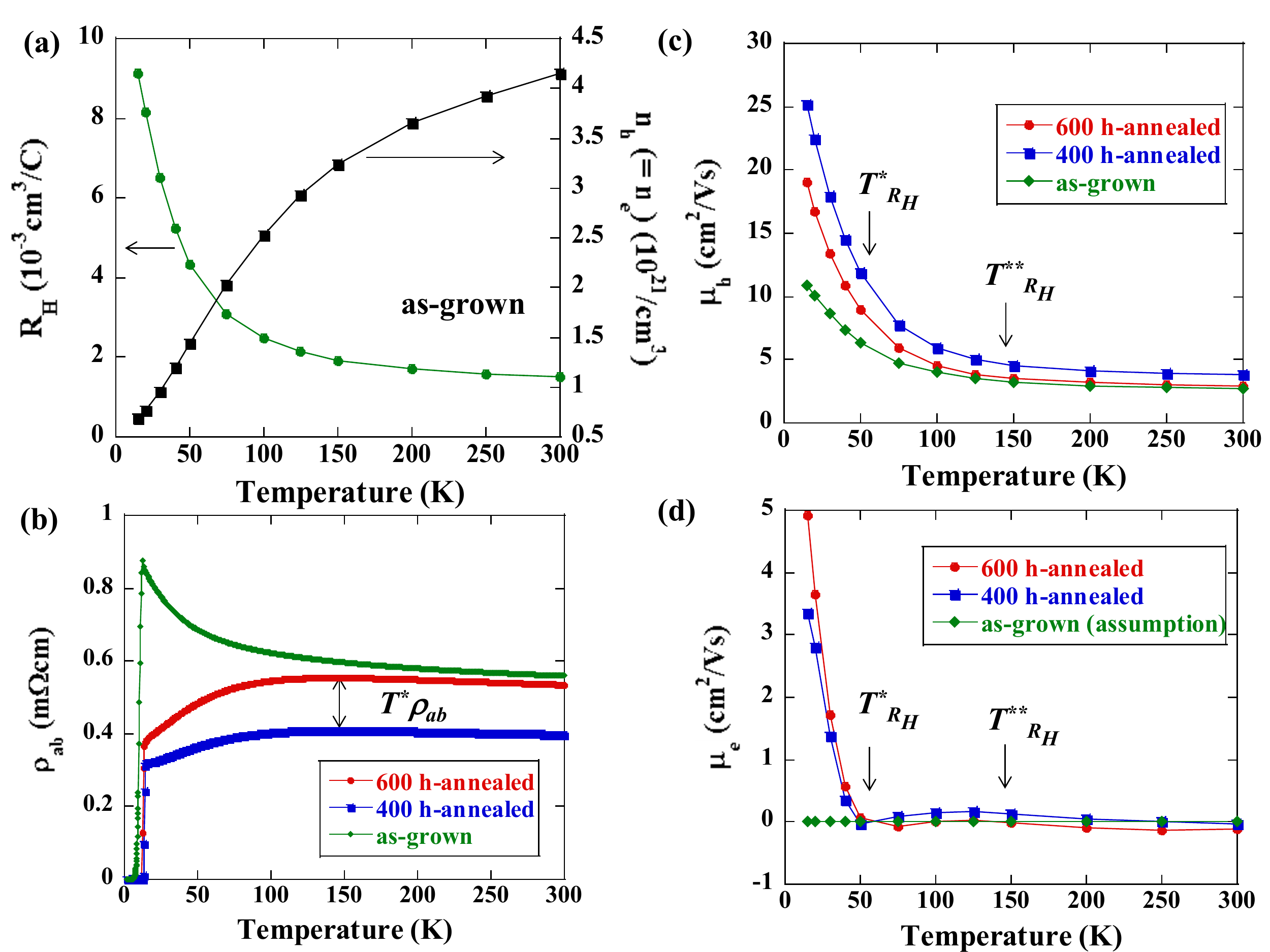}
			\caption{\label{fig4}(Color online) (a) Temperature dependence of Hall coefficient, $R_{H}$, and the hole number, $n$, for the as-grown Fe$_{1+y}$Te$_{0.8}$Se$_{0.2}$ single crystal. (b) In-plane resistivities, $\rho_{ab}(T)$, for the as-grown, 400 h-annealed, and 600 h-annealed Fe$_{1+y}$Te$_{0.8}$Se$_{0.2}$ single crystals. (c) Temperature dependence of hole mobilities, $\mu_{h}$, for the as-grown, 400 h-annealed, and 600 h-annealed Fe$_{1+y}$Te$_{0.8}$Se$_{0.2}$ single crystals. (d) Temperature dependence of electron mobilities, $\mu_{e}$, for the as-grown, 400 h-annealed, and 600 h-annealed Fe$_{1+y}$Te$_{0.8}$Se$_{0.2}$ single crystals. The arrows indicate the characteristic temperatures, $T^*_{\rho_{ab}}$, $T^{**}_{R_{H}}$, and $T^*_{R_{H}}$.}
		\end{center}
	\end{figure}
	
	
	Additional insight into the anomalous $R_{H}$ was obtained by measuring the magnetoresistance (MR) for these crystals (x = 0.2) at several temperatures above $T_c$ (Fig. \ref{fig3}(a)-\ref{fig3}(c)). In all cases, negative MR is observed, which is in contrast to the large positive MR for an O$_2$ annealed x = 0.4 single crystal~\cite{sun}, but agrees with the results obtained for polycrystals of x = 0.1 and 0.2~\cite{tropeano}. Because the MR exhibits $H^2$ behavior, the coefficient $A$ is plotted as a function of the temperature in Fig. \ref{fig3}(d). In all cases, at higher temperatures, $A$ is very small and temperature independent; however, the magnitude of $A$ gradually increases below $T^{**}_{R_{H}}$ and rapidly increases below $T^*_{R_{H}}$. Overall, the behavior of these samples is similar except for the fact that the increase in $A$ below $T^*_{R_{H}}$ seems faster for the annealed samples than for the as-grown sample. To clarify the difference, we plotted MR as a function of $\tan\Theta_{H} \equiv \sigma_{xy}/\sigma_{xx}$ at several temperatures for these samples in Fig. \ref{figS2}(a)-\ref{figS2}(c). In the as-grown sample, all the MR data lie on one curve, indicating that the modified Kohler's rule ($\Delta\rho/\rho(0) \propto \tan^2\Theta_{H}$, but MR is \textit{negative} here) and this holds for the entire temperature and magnetic field range that was measured (Fig. \ref{figS2}(a)). On the other hand, in the Te-annealed samples, the MR data above $T^*_{R_{H}}$ ($\approx$ 50 K) fall on one curve. The reason why the modified Kohler's rule holds for negative MR is currently unknown; however, these results may be the reflection that, in the as-grown sample, transport is effectively dominated by one hole band, whereas in the Te-annealed samples, a phase of a multi-band nature appears below 50 K. Consequently, the behavior of MR corresponds well with the temperature dependence of $R_{H}$. 
	
	The large $R_{H}$ and related strong temperature dependence have frequently been observed in Fe-based high-$T_c$ superconductors~\cite{tsukada,tamegai,sun,kasahara,liu,tropeano,eom,sun1,albenque,fang1} including Fe$_{1+y}$Te$_{1-x}$Se$_{x}$. The origin of this behavior has been discussed in terms of anisotropic carrier scattering due to strong AFM fluctuations~\cite{kasahara,liu,eom} or multiband effects with electron-hole asymmetry~\cite{tamegai,sun,sun1}. The former interpretation was first successfully applied to high-$T_c$ cuprates or heavy fermion systems~\cite{kontani}. However, if the electron transport is dominated by either anisotropic carrier scattering or multiband effects, the orbital MR should be positive. Therefore, we would have to search for another cause to interpret the observed negative MR. One possibility may be the Kondo effect due to magnetic impurity scattering. Here, the excess iron, which is inevitably incorporated in the as-grown crystal~\cite{koshika}, could be the magnetic impurity responsible for the negative MR~\cite{tropeano}. Indeed, our as-grown crystals show semiconducting temperature dependence of $\rho_{ab}$ (Fig. \ref{fig1}(a) and Fig. \ref{fig4}(b)), which is the typical behavior of $\rho_{ab}$ associated with the existence of a magnetic impurity. However, we observe negative MR (Fig. \ref{fig3}(b) and (c)) even when the excess iron is removed and $\rho_{ab}$ shows metallic behavior (Fig. \ref{fig1}(b) and Fig. \ref{fig4}(b)). Therefore, at least for the annealed crystals, this interpretation to explain the negative MR is unlikely. Another possibility is that the negative MR is caused by the pseudogap effect. High-$T_c$ cuprates are well known to exhibit negative out-of-plane MR due to the recovery of the electronic density-of-states (DOS) along with the suppression of the pseudogap under magnetic fields~\cite{t.shiba, wata3}. In our work, we may have observed a similar effect for $\rho_{ab}$. Very recently, a pseudogap opening below 150 K was directly observed by ARPES in the electron band around the M point for a Te-annealed sample with $x$ = 0.2~\cite{koshiishi}.     
	
	We examined what happens at $T^{**}_{R_{H}}$ or $T^{*}_{R_{H}}$ by analyzing the observed transport coefficients, $\rho_{ab}$ and $R_{H}$, of samples with the same Se concentrations ($x$ = 0.2). For this purpose, we adopt a simple two-band model with an equal number $n$ of electrons and holes, assuming FeTe$_{1-x}$Se$_{x}$ is a compensated semimetal, but permitting $n$ to be temperature dependent. This assumption is justified in two ways. The first is our observation of the field-linear behavior in $\rho_{yx}$ (Fig. 3(a)-(c)). In a two-band model, the Hall coefficient $R_{H}$ becomes field dependent when the number of holes and electrons are different~\cite{sun}. However, only when they are equal, $R_{H}$ takes a constant value, resulting in $\rho_{yx}$ becoming linear in the magnetic field. Therefore, the observed field-linear $\rho_{yx}$ indicates that the number of holes and electrons are the same (or nearly equal). The second justification is the estimation of the number of holes and electrons from the Fermi surface area by using ARPES~\cite{koshiishi1}. For the sample with $x$ = 0.2, the number of holes and electrons are estimated to be 0.304 and 0.300 /unit cell, respectively, which assures that they are almost the same. 
	
	Then, the in-plane resistivity and Hall coefficient are described as, $\rho_{ab} = \frac{1}{ne(\mu_{e}+\mu_{h})}$ and $R_{H} = \frac{\mu_{h}-\mu_{e}}{ne(\mu_{e}+\mu_{h})}$, respectively, where $\mu_{e}$ ($\mu_{h}$) is the mobility of electrons (holes). First, $\mu_{e}$ of the as-grown sample is assumed to be zero, because $R_{H}$ of this sample is always positive and smoothly evolves with temperature (Fig. \ref{fig2}(a) and (d)), which implies that the electron contribution to the transport properties is small. Thus, in the as-grown sample, the two-band model effectively results in a one-band model. Then, $n$ can be estimated using the observed $R_{H}$ as $n = \frac{1}{eR_{H}}$ (Fig. \ref{fig4}(a)). The number of carriers (holes) $n$ gradually decreases as the temperature decreases from room temperature to $\approx$ 150 K, but it rapidly decreases consistent with the assumption that the pseudogap opens below this temperature. Figure \ref{fig4}(b) shows $\rho_{ab}$ for the the as-grown, 400 h-annealed, and 600 h-annealed samples. One may notice that, at temperatures below $\approx$ 150 K, $\rho_{ab}$ of the as-grown sample gradually increases as the temperature decreases, whereas $\rho_{ab}$ of the annealed samples show metallic behavior below this temperature. Next, for each of the samples, $\mu_{h}$ and $\mu_{e}$ are estimated using $\rho_{ab}$ (Fig. \ref{fig4}(b)) and $R_{H}$ (Fig. \ref{fig2}(d)) with the two-band model assuming that the obtained $n$ is common for all the samples. A small amount of excess Fe may not alter the overall band structure (i.e., the number of carriers $n$). The results are plotted in Fig. \ref{fig4}(c) and (d), respectively. It is evident that the values of $\mu_{h}$ of the annealed samples increase much faster than those of the as-grown sample below 150 K ($\approx$ $T^{**}_{R_{H}}$). In addition, as expected considering the temperature dependence of $R_{H}$ (Fig. \ref{fig2}(d)), the values of $\mu_{e}$ of the annealed samples appear below $T^{*}_{R_{H}}$ ($\approx$ 50 K). These results indicate that, in the as-grown sample, $\rho_{ab}$ is semiconducting ($d\rho_{ab}/dT \textless$ 0) primarily owing to the decrease in $n$ below $T^{**}_{R_{H}}$, whereas in the annealed samples, $\rho_{ab}$ is metallic ($d\rho_{ab}/dT \textgreater$ 0) below this temperature. This is because the increase in $\mu_{h}$ surpasses the decrease in $n$ below $T^{**}_{R_{H}}$. In addition, below $T^{*}_{R_{H}}$, the appearance of $\mu_{e}$ for the annealed samples further contributes to in-plane metallic conduction. The appearance of electron carriers is consistent with a recent ARPES measurement in which clear electron pockets around the M point, as well as hole pockets around the $\Gamma$ point, were observed for Te-annealed samples with x = 0.2~\cite{koshiishi}. The increase in the mobilities implies an enhancement of the carrier lifetime $\tau$, thus it implies that the electronic states become coherent. We further confirmed this kind of incoherent to coherent crossover transition by ARPES measurements~\cite{koshiishi1}. Consequently, in the superconducting samples, the electron bands become coherent below $T^{*}_{R_{H}}$, whereas the hole bands become coherent below $T^{**}_{R_{H}}$.
	
	\begin{figure}[t]
		\begin{center}
			\includegraphics[width=85mm]{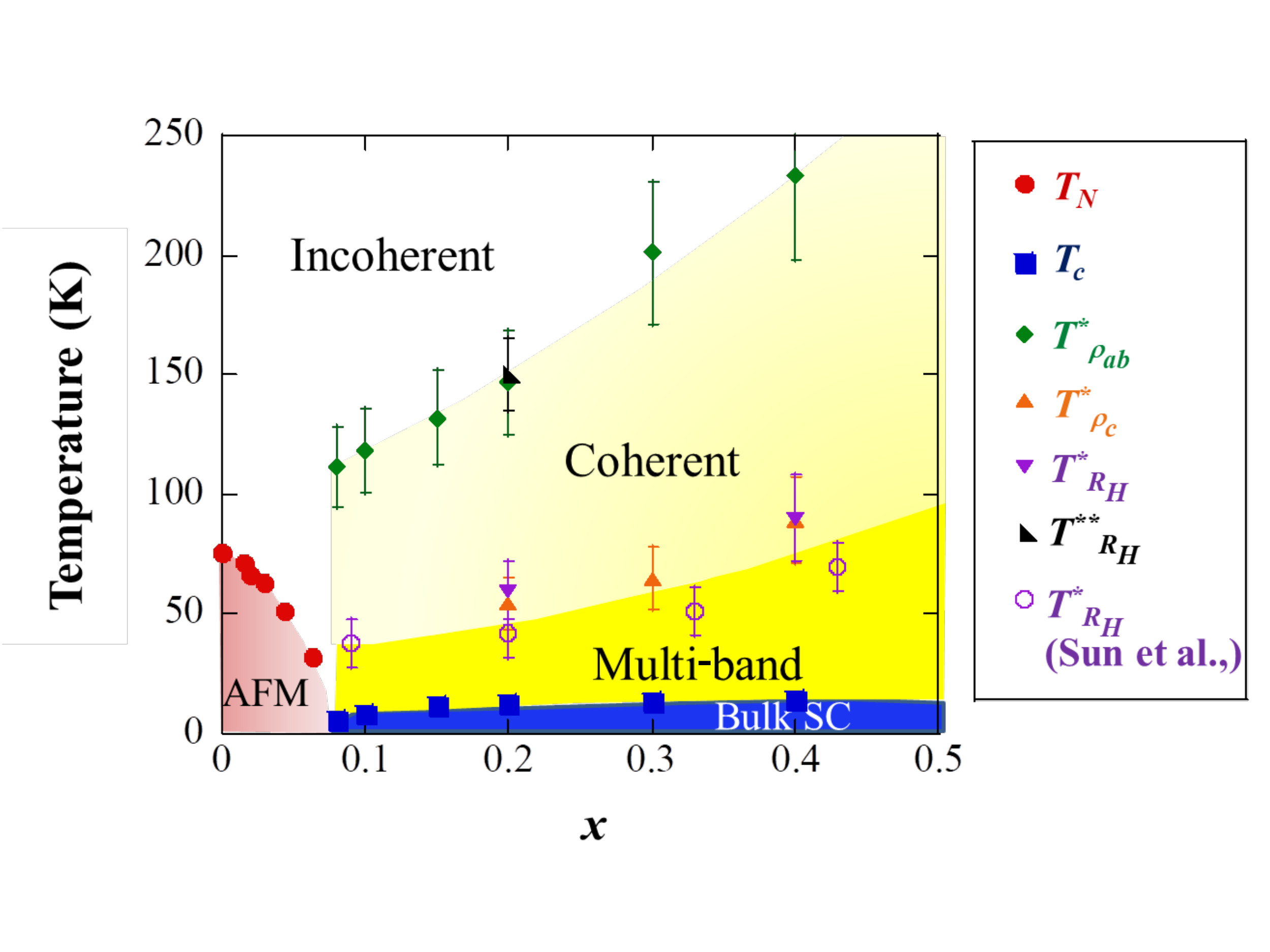}
			\caption{\label{fig6}(Color online) Characteristic temperatures vs. Se concentration $x$ for Te-annealed Fe$_{1+y}$Te$_{1-x}$Se$_{x}$. Closed symbols represent the characteristic temperatures obtained in this study. Open circles of $T^*_{R_{H}}$ are replotted from ref. ~\cite{tamegai}.}
		\end{center}
	\end{figure}

	The characteristic temperatures $T^*_{\rho_{ab}}$, $T^*_{\rho_{c}}$, $T^{**}_{R_{H}}$, $T^*_{R_{H}}$, $T_{N}$, and $T_{c}$ are plotted as a function of the Se concentration $x$ in Fig. \ref{fig6}. The long-range AFM ordered state exists only in the range of 0 $\le$ $x$ $\le$ 0.06 and the superconducting state emerges at $x$ $\ge$ 0.08. The fact that the AFM and superconducting states do not coexist agrees with the recent report on Fe$_{1+y}$Te$_{1-x}$Se$_{x}$~\cite{tamegai}, which implies that the static AFM order competes with superconductivity in this system. However, a comparison of their results~\cite{tamegai} with ours reveals some differences. First, the phase boundary is slightly different; it exists at approximately $x$ = 0.07 in this study, whereas in the other study it is asserted as being at x = 0.05~\cite{tamegai}. Second, more importantly, the AFM state is gradually suppressed with increasing $x$, and finally it is replaced by superconductivity at approximately $x$ = 0.07 as in a quantum critical phase transition, in contrast to the sudden (first-order-like) transition observed previously~\cite{tamegai}. These differences may have their origins in the different annealing procedures that were used. Our result on the relation between the AFM and superconducting states is similar to the phase diagram of CeFeAsO$_{1-x}$F$_x$~\cite{zhao} or heavily doped LaFeAsO$_{1-x}$H$_x$~\cite{sakurai}, but appears to be different from that obtained for LaO$_{1-x}$F$_x$FeAs~\cite{luetkens}. Our phase diagram is apparently different from that of Ba(Fe$_{1-x}$Co$_x$)$_2$As$_2$~\cite{nandi} where AFM and superconductivity coexist. 
	
	For the superconducting samples with $x$ $\ge$ 0.08, the multi-band nature always appears before superconductivity sets in. This result is consistent with the scenario where pairing is mediated by AFM fluctuations via Fermi surface nesting between the holes and electron Fermi pockets~\cite{mazin,kuroki}. Moreover, it should be remarked that the incoherent to coherent crossover transition always appears before the multi-band nature or superconductivity sets in.
	
	As noted above, $T^{**}_{R_{H}}$ and $T^*_{\rho_{ab}}$ coincide. This implies that opening of the pseudogap is correlated with the coherent transition. Moreover, $T^*_{R_{H}}$ and $T^*_{\rho_{c}}$ coincide. Because the participation of electron carriers in charge transport is the cause for $T^*_{R_{H}}$, this would also be expected to be the cause for anomalous $T^*_{\rho_{c}}$. In the annealed samples, $\rho_{ab}$ is metallic but $\rho_{c}$ is semiconducting in the temperature range between $T^*_{\rho_{ab}}$ and $T^*_{\rho_{c}}$. This is very similar to the behavior of the cuprates~\cite{usu}, for which the Fermi surface is two-dimensional (2D) and the out-of-plane hopping probability, $t_{\perp}$, reaches its maximum at the anti-node where the pseudogap opens~\cite{ioffe}. Thus, we assume that the hole band in the Fe$_{1+y}$Te$_{1-x}$Se$_{x}$ system is similar to that of the cuprates. On the other hand, when the electron band appears below $T^*_{\rho_{c}}$, $\rho_{c}$ shows a plateau (Fig. \ref{fig1} (d)). This is because the out-of-plane conductivity originated in the electron band is added to the semiconducting conductivity of the hole bands. This indicates that the out-of-plane conductivity of the electron band is rather metallic. On the basis of this result, we suppose that $t_{\perp}$ of the electron band does not depend much on the in-plane wavenumber when the band is 2D, or the electron band is anisotropically three-dimensional (3D). Here, we would like to emphasize that these temperatures ($T^*_{\rho_{ab}}$, $T^*_{\rho_{c}}$, and $T^*_{R_{H}}$) appear only in the superconducting (annealed) samples. It should be noted that $T^*_{\rho_{ab}}$ cannot be recognized in the samples with low Se concentrations (x $\le$ 0.06) and those samples are not superconducting, irrespective as to whether they are annealed (Fig. \ref{fig1} (a) and (b)). 

	Based on this analysis, we conclude that superconductivity emerges from the coherent electronic state that is accompanied by opening of the pseudogap in this Fe$_{1+y}$Te$_{1-x}$Se$_{x}$ system. In other words, the existence of the coherent state with the pseudogap is a prerequisite for the occurrence of superconductivity. The importance of the coherent electronic state for the emergence of superconductivity has been inferred by ARPES measurements~\cite{ieki,shen}. Here, we verified, using transport measurements, that the incoherent to coherent transition and opening of the pseudogap always take place in the superconducting Fe$_{1+y}$Te$_{1-x}$Se$_{x}$ samples at the doping($x$)-dependent characteristic temperature $T^{**}_{R_{H}}$ (or $T^*_{\rho_{ab}}$) above $T_{c}$.     
	
	\section{IV. DISCUSSION}		
	\def\Vec#1{\mbox{\boldmath $#1$}}
	
	We next discuss the origin of the pseudogap. The observation of the pseudogap has been reported for other Fe-based superconductors such as Ba$_{0.75}$K$_{0.25}$Fe$_{2}$As$_{2}$~\cite{xu}, BaFe$_{2}$(As$_{1-x}$P$_{x}$)$_{2}$~\cite{shimojima}, and FeSe~\cite{rossler}, and the origin thereof has been explained in several ways. Here, in the Fe$_{1+y}$Te$_{1-x}$Se$_{x}$ system, we observed the transformation of the electronic state to occur in two steps: at higher temperature $T^{**}_{R_{H}}$ (or $T^*_{\rho_{ab}}$) and at lower temperature $T^{*}_{R_{H}}$ (or $T^*_{\rho_{c}}$). We suppose that the pseudogap opens at both of the temperatures $T^{**}_{R_{H}}$ and $T^{*}_{R_{H}}$; the former (latter) corresponds to the temperature below which the pseudogap gradually (substantially) opens. One possibility for the occurrence of the pseudogap is the fluctuations of the electronic nematic orders. First, we consider the origin of the pseudogap at $T^{**}_{R_{H}}$. In strongly correlated iron chalcogenides such as Fe$_{1+y}$Te$_{1-x}$Se$_{x}$, an orbital-selective Mott phase (OSMP) and related incoherent to coherent transition has been proposed to exist theoretically~\cite{yu, yin1} and it has actually been observed by ARPES~\cite{shen, shen1}, in which only the $d_{xy}$ orbital state (coherent at lower temperatures) loses spectral weight at higher temperatures due to strong on-site Coulombic interactions $U$ and Hund's coupling $J$. Thus, our observation for the coherent transition below $T^*_{\rho_{ab}}$ may correspond to the transformation from OSMP to the metallic state. On the basis of the experimental observation (Fig. \ref{fig4} (c)), we assume that the hole bands become coherent below this temperature. We consider this coherent transition to be responsible for triggering the opening of the pseudogap. When the states become coherent, the Fermi surfaces become well defined with some kind of band hybridization, which would cause the pseudogap to open through inter-band nesting. Simultaneously, this inter-band nesting would enhance the orbital nematic fluctuations~\cite{kontani1}. In fact, elastoresistance measurements have revealed a strong nematic response for FeTe$_{0.6}$Se$_{0.4}$ below $T_{s}$ ($\approx$ $T^{**}_{R_{H}}$ or $T^*_{\rho_{ab}}$), similar to other optimally doped Fe-based superconductors~\cite{kuo}. Furthermore, energy splitting between the $d_{xz}$ and $d_{yz}$ bands at the $\Gamma$ point, which is evidential of orbital fluctuations, has been observed at low temperatures~\cite{johnson}. An elastic constant $C_{66}$ in this compound showed substantial ($\approx$ 40 \%) softening ~\cite{yoshizawa2} from high temperatures corresponding to $T^{**}_{R_{H}}$ in a similar manner to Ba(Fe$_{1-x}$Co$_x$)$_2$As$_2$~\cite{yoshizawa1}. The electron-lattice coupling constant $\lambda$ evaluated by Jahn-Teller analysis was large ($\approx$ 0.15 eV/Fe), further supporting the scenario for the development of the orbital nematic fluctuations.
	
	Next, we consider the origin of the pseudogap at $T^{*}_{R_{H}}$. Considering the experimental observation (Fig. \ref{fig4} (d)), we assume that the electron bands also become coherent below this temperature as part of the coherent transition originated in the OSMP. Then, this coherent transition strengthens the inter-band nesting with additional band hybridization, which would result in substantial pseudogap opening. This improved nesting would enhance the spin nematic fluctuation as well~\cite{fernandes}. Actually, in the optimally superconducting FeTe$_{1-x}$Se$_{x}$, the ratio of the strength of spin correlations with a single-stripe AFM wave vector \Vec{Q}$_{SAF}$ = ($\pi$, 0) (here we use \Vec{Q}-vector notation in the 1-Fe unit cell) that connects the hole and electron pockets to that with a double-stripe AFM wave vector \Vec{Q}$_{DSAF}$ = ($\pi$/2, $\pi$/2) has been shown to grow~\cite{z.xu} similar to the nematic response of the elastoresistance measurements~\cite{kuo}. This result suggests that the single-stripe AFM fluctuations are responsible for the nematicity especially at low temperatures below $T^{*}_{R_{H}}$. We consider that \Vec{Q}$_{DSAF}$ = ($\pi$/2, $\pi$/2), which developed in the parent compound Fe$_{1+y}$Te, competes with  \Vec{Q}$_{SAF}$ = ($\pi$, 0) causing $T^{*}_{R_{H}}$ to decrease with decreasing $x$, and eventually to disappear at $x$ = 0.07.
	
	The validity of this assumption was investigated by elastic constant measurements. The elastic constant $C_{44}$ in this compound showed softening at low temperatures, but the amount of softening was small ($\approx$ 5 \%), and the $\lambda$ was small ($\approx$ 0.03 eV/Fe)~\cite{yoshizawa} compared with $\lambda$ (0.22 - 0.25 eV/Fe) for $C_{66}$ of Ba(Fe$_{1-x}$Co$_x$)$_2$As$_2$~\cite{yoshizawa1}. This result implies that the development of spin correlation with \Vec{Q}$_{DSAF}$ = ($\pi$/2, $\pi$/2) is responsible for the anomalous $C_{44}$~\cite{yoshizawa}. On the other hand, the development of \Vec{Q}$_{SAF}$ = ($\pi$, 0) correlation may contribute to the anomalous $C_{66}$. With increasing Te-annealing durations, the $C_{44}$ anomaly decreased, whereas the $C_{66}$ anomaly increased~\cite{yoshizawa2}. This result implies that the spin correlation with \Vec{Q}$_{DSAF}$ = ($\pi$/2, $\pi$/2) was weakened whereas that with \Vec{Q}$_{SAF}$ = ($\pi$, 0) was strengthened by the annealing. This is consistent with the assumption that double-stripe AFM order competes with single-stripe AFM fluctuation, and thus supports the scenario that the spin nematic fluctuation based on the \Vec{Q}$_{SAF}$ = ($\pi$, 0) correlation is responsible for the pseudogap opening below $T^{*}_{R_{H}}$. Details will be published elsewhere.

	Even in the as-grown sample, we assume that some kind of pseudogap opens below $\approx$ 150 K. However, the origin may be different from that of the superconducting samples mentioned above. Excess iron in the as-grown samples may suppress the electronic coherence and thus the spin correlations with \Vec{Q}$_{SAF}$ = ($\pi$, 0); instead, it may stabilize spin correlations of the parent compound, \Vec{Q}$_{DSAF}$ = ($\pi$/2, $\pi$/2), causing another pseudogap to open. In this case, $T^{*}_{R_{H}}$ never appear, because $T^{*}_{R_{H}}$ is associated with the development of the spin correlations with \Vec{Q}$_{SAF}$ = ($\pi$, 0). In fact, development of the spin correlations with \Vec{Q}$_{DSAF}$ = ($\pi$/2, $\pi$/2) have been reported in non-superconducting FeTe$_{1-x}$Se$_{x}$ with $x$ = 0.45 and excess iron~\cite{z.xu}.        
	
	Another possible explanation for the existence of the pseudogap below $T^{*}_{R_{H}}$ is preformed Cooper pairing. Recently, Fe-based superconductors, especially the ``11" system, have been argued to exist deep inside the crossover regime between weak-coupling Bardeen-Cooper-Schrieffer (BCS) and strong-coupling Bose-Einstein condensate (BEC). In FeSe, the onset of strong nonlinear diamagnetism, which provides evidence for the prevailing phase fluctuations of superconductivity, has been shown at $\approx$ 20 K~\cite{kasahara1}, which more than twice exceeds its $T_c \approx$ 8.5 K. FeTe$_{1-x}$Se$_{x}$ has been shown to exhibit a large superconducting gap $\Delta$ and small values for the Fermi energy $\epsilon_F$~\cite{Lubashevsky,Okazaki}; the ratio $\Delta/\epsilon_F \approx$ 0.5 suggests that this system is within the BCS-BEC crossover regime. Therefore, preformed Cooper pairing above $T_c$ may be possible in this FeTe$_{1-x}$Se$_{x}$ system. Much more effort would be required to clarify the origin of the pseudogap and to understand the role it plays in superconductivity.     

	In summary, the $x$-$T$ phase diagram of Fe$_{1+y}$Te$_{1-x}$Se$_{x}$ (0 $\le x \le$ 0.4) was studied by performing various transport measurements on samples with and without Te-annealing. In the superconducting samples (Te-annealed samples with 0.08 $\le x \le$ 0.4), the incoherent to coherent crossover transition, which is accompanied by the opening of the pseudogap, was found to occur in two steps. First, only the in-plane state becomes coherent at higher temperatures below $T^{**}_{R_{H}}$ (and $T^*_{\rho_{ab}}$ $\approx$ 150 K for $x$ = 0.2). Second, both the in-plane and out-of-plane states become coherent with the multi-band nature at lower temperatures below $T^*_{R_{H}}$ (and $T^*_{\rho_{c}}$ $\approx$ 50 K for $x$ = 0.2). Based on these results, we established a new $x$-$T$ phase diagram, which may be inherent to the strongly correlated Fe$_{1+y}$Te$_{1-x}$Se$_{x}$ system. The phase diagram reveals that, not only the emergence of the phase with a multi-band nature but also the coherent transition with the pseudogap, are primarily important for the occurrence of superconductivity.      

	\section{Acknowledgments}
	\begin{acknowledgments}
		
		We thank K. Koshiishi, A. Fujimori, H. Kontani, S. Onari, T. Tamegai, and Y. Koike for their helpful discussions. The magnetotransport measurements
		were mostly performed using PPMS at Iwate University. Some of the measurements were performed at the High Field Laboratory for Superconducting Materials, Institute for Materials Research, Tohoku University (Project No. 14H0007). 
	\end{acknowledgments}

\end{document}